\documentclass[aps,prl,reprint]{revtex4-1}

\usepackage{graphicx}% Include figure files

\draft % marks overfull lines with a black rule on the right

\begin{document}
% Use the \preprint command to place your local institutional report number 
% on the title page in preprint mode.
% Multiple \preprint commands are allowed.
%\preprint{}

\title{Fast nonthermal processes in pulsed laser deposition} %Title of paper
%\title{Homoepitaxial growth of SrTiO$_3$ by Pulsed Laser Deposition: energetic vs thermal growth} %Title of paper

% repeat the \author .. \affiliation  etc. as needed
% \email, \thanks, \homepage, \altaffiliation all apply to the current author.
% Explanatory text should go in the []'s, 
% actual e-mail address or url should go in the {}'s for \email and \homepage.
% Please use the appropriate macro for the type of information

% \affiliation command applies to all authors since the last \affiliation command. 
% The \affiliation command should follow the other information.

\author{Jeffrey G. Ulbrandt}
\author{Xiaozhi Zhang}
\affiliation{Department of Physics and Materials Science Program, University of Vermont,  Burlington VT 05405}
\author{Rui Liu}
\affiliation{Department of Physics and Astronomy, Stony Brook University, Stony Brook NY 11794}
\author{Kenneth Evans-Lutterodt}
\affiliation{National Synchrotron Light Source II, Upton NY 11967}
\author{Matthew Dawber}
\affiliation{Department of Physics and Astronomy, Stony Brook University, Stony Brook NY 11794}
\author{Randall L. Headrick}
\email[]{rheadrick@uvm.edu}
\affiliation{Department of Physics and Materials Science Program, University of Vermont, Burlington VT 05405}

% Collaboration name, if desired (requires use of superscriptaddress option in \documentclass). 
% \noaffiliation is required (may also be used with the \author command).
%\collaboration{}
%\noaffiliation

\date{\today}

%  SrTiO$_3$ (STO)

\begin{abstract}

Pulsed Laser Deposition (PLD) is widely used to grow epitaxial thin films of quantum materials such as complex oxides.   Here, we use in-situ X-ray scattering to study homoepitaxy of SrTiO$_3$  by energetic (e-) and thermalized (th-) PLD.  We find that e-PLD suppresses the lateral growth of two-dimensional islands, which suggests that energetic particles break up smaller islands.  Fast interlayer transport occurs for both e-PLD and th-PLD, implying a process operating on sub-microsecond timescales that doesn't depend strongly on the kinetic energy of the incident particles.

\end{abstract}

% insert suggested PACS numbers in braces on next line

\maketitle %\maketitle must follow title, authors, abstract and \pacs

% Main Text

Pulsed Laser Deposition (PLD) is a versatile method for homoepitaxial growth of ultra-smooth crystalline surfaces,\cite{Vasco_2008} heteroepitaxial interfaces,\cite{Ohtomo_2004, Reyren_2007,Caviglia_2008} and superlattices.\cite{Lee_2005}  Understanding and controlling the growth processes occurring in PLD and other energetic deposition methods presents an important challenge for the synthesis of thin layers with controllable properties. 

PLD employs microsecond-scale pulses that occur as the laser plume reaches the growth surface, resulting in instantaneous deposition rates  orders of magnitude higher than in continuous deposition methods such as  Molecular beam Epitaxy (MBE). Deposited particles typically cannot diffuse  a significant lateral distance on the surface while the rest of the plume is arriving.   This is generally true over a wide range of deposition temperatures in the layer-by-layer growth regime, which is  most relevant to epitaxial growth. Thus, it becomes possible to study fast microscopic crystallization mechanisms separately from slower competing thermal relaxation effects by studying the evolving X-ray scattering intensity on the relevant timescales.

% Figure 1 - Experimental Set-up
\begin{figure}[ht!]
\includegraphics[scale=0.5]{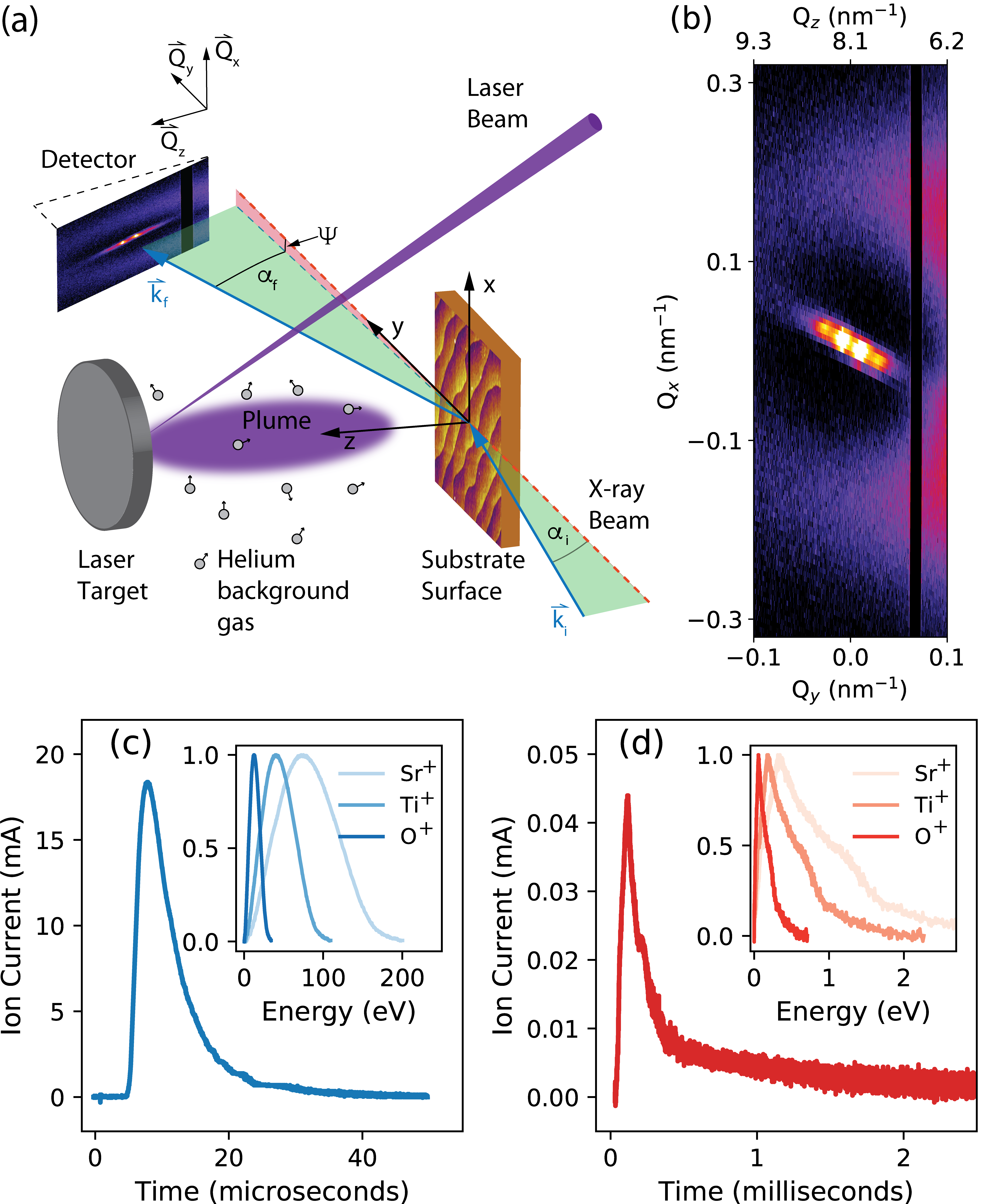}
\caption{\label{Experiment} (a) Schematic of the X-ray scattering experiment. (b) A portion of the scattering pattern transformed into Q-coordinates.  (c) Time-of-flight (ToF) spectra for ions in the e-PLD process. The inset shows the corresponding energy distribution for the ionic species.  (d) The corresponding ToF for th-PLD.}
\end{figure}

% Figure 2 - Specular and Diffuse Comparison
\begin{figure*}[ht!]
\includegraphics[scale=0.5]{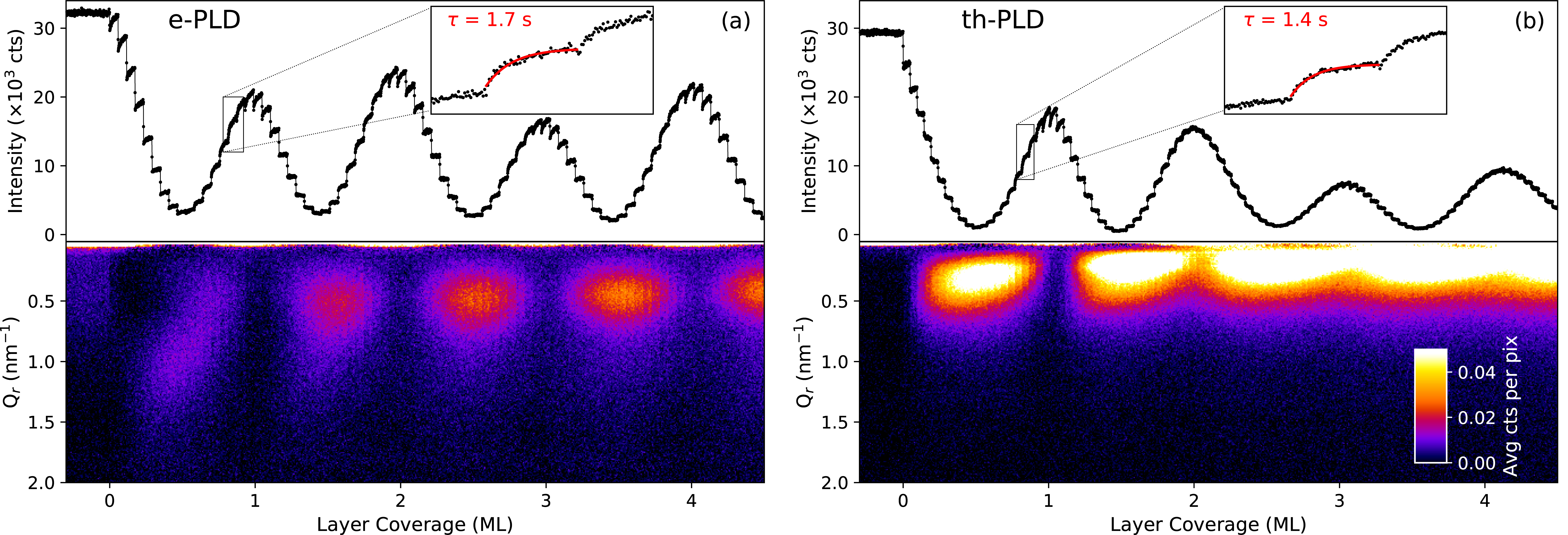}
\caption{\label{Spec_Diff} Total specular and diffuse scattering comparison between the energetic process (a) and the thermalized process (b) for a 6 second dwell time between pulses. The top panels in (a) and (b)  show the evolution of the total specular intensity. and the bottom panes show the corresponding diffuse scattering as a function of Q$_r$.  The insets in (a) and (b) show the thermal recovery of the specular intensity for a single pulse near 0.8 ML coverage. Note that the time axis has been converted to layer coverage for convenience, although coverage increases in discrete steps at each laser pulse.}
\end{figure*}

Previous research has focused on the possibility that arriving particles with hyperthermal kinetic energies in the range of tens to hundreds of electron volts typical of PLD produce energetic surface smoothening effects, including energetic-particle island breakup and non-thermal transient enhanced mobility.  Transient enhanced mobility is thought to involve conversion of hyperthermal kinetic energy into ballistic motion of deposited species.\cite{Fleet_2005, Tischler_2006,Vasco_2007} The motion can be in-plane -- like enhanced surface diffusion  -- or it can involve material transferring between layers, such as when material lands on top on a unit-cell height island and then hops over the step edge to the next lower terrace.  For example, in an experimental study of SrTiO$_3$ homoepitaxy by Tischler et al., it was observed that inter-layer transport occurs on two different time scales separated by orders of magnitude.\cite{Tischler_2006}    Another study by Fleet et al. deduced the presence of a fast inter-layer relaxation process by comparing the experimentally-determined X-ray scattering intensity change immediately after the deposition pulse to the theoretical change for random deposition without relaxation.\cite{Fleet_2005}

An additional energetic mechanism was suggested by Willmott et al.\cite{Willmott_2006} for PLD growth of La$_{1-x}$Sr$_x$MnO$_3$ on SrTiO$_3$, in which the energetic species in the laser plume breaks up islands into smaller daughter islands. Island break-up is expected to delay the average island size from reaching the critical size at which the next layer begins to nucleate.  This delay should lead to a relatively smoother growth surface.  These results seem to confirm computational predictions of  a break-up effect  in which adatoms are ``chipped" from the edges of larger islands.\cite{Pomeroy_2002}  Supporting evidence has also been found in an experimental study of Platinum homoepitaxy by PLD, in which deposited particles with energies above 200 eV result in a higher island nucleation density,  attributed to an increase of adatoms pushed out of the surface by the impinging energetic particles.\cite{Schmid_2009}  On the other hand, later  studies of SrTiO$_3$ homoepitaxy utilizing X-ray scattering measurements sensitive to in-plane structure have shown that the surface in-plane length scale coarsens significantly during the growth.\cite{Eres_2016, Ferguson_2009} This was interpreted as evidence that significant island breakup does not occur.

As we have mentioned above, we define the PLD regime as being the case where all the material in the laser plume is deposited before it diffuses significantly or nucleates islands, so that in the absence of fast non-thermal processes the instantaneous pulse (impulse) approximation is valid. This has important implications for the length scale of the islands, particularly during the initial stages of recovery after a deposition pulse.  In continuous deposition the peak island density scales  as $F^\chi$ where $F$ is the deposition flux and $\chi$ is an exponent that depends on the size of the critical nucleus.\cite{Venables_1984} On the other hand, in PLD the nucleation density in the impulse approximation is set by the pulse intensity $\sigma$, with a length scale of $\ell \approx a/\sigma^{1/2} $ where $a$ is the lattice constant.  This is strictly only true for the first deposited pulse, while for later pulses there will be a distribution of islands from previous pulses that have  grown larger by aggregation and ripening in addition to islands from the most recent deposition pulse. 

%We note that Hinneman et al. have shown that this process leads to a novel scaling relationship for the island size distribution in the submonolayer regime as a function of the pulse intensity and the deposition time.\cite{Hinnemann_2001}

% Figure 3 - Diffuse Scattering Comparison
\begin{figure*}[ht!]
\includegraphics[scale=0.65]{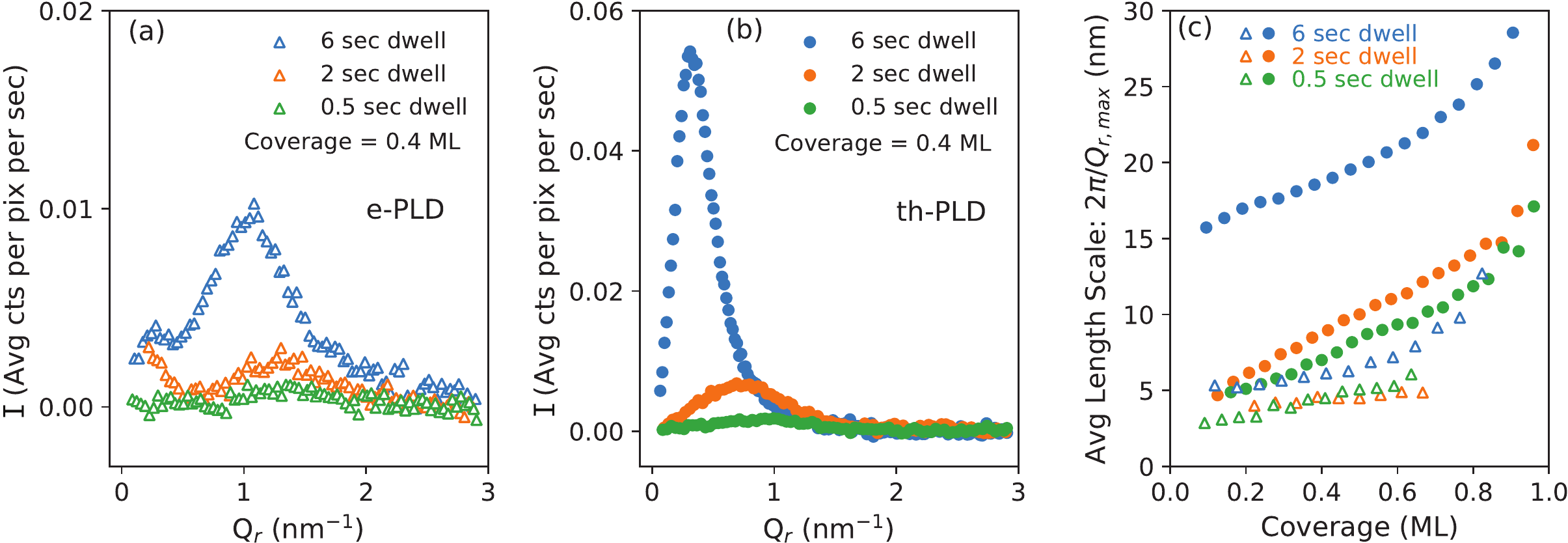}
\caption{\label{Dwell}Diffuse scattering profiles for (a) the energetic process, and (b) the thermalized process at 0.4 ML coverage. (c) Average length scale during first monolayer deposition. The triangles are for the e-PLD and the circles are for the th-PLD.}
\end{figure*}

Here, we present in-situ X-ray scattering measurements of  homoepitaxial growth of SrTiO$_3$, by comparing the standard energetic deposition process (e-PLD) with a thermalized version of PLD (th-PLD).   We report two principal results: (i)  The length scale is reduced and the nucleation density is greatly increased in e-PLD relative to th-PLD. We observe for a pulse intensity $\sigma \approx 1/20$ ML, the in-plane length scale $\ell$ in  e-PLD can approach the limiting value deduced above, which for $a=0.3905$ nm is $\approx$2 nm.    Coarsening is also limited during e-PLD, consistent with island breakup.  (ii) We find that fast interlayer transport occurs during both e-PLD and th-PLD.  This result is remarkable since by result (i)  th-PLD  also produces larger island sizes, which implies that deposited particles have to migrate significantly longer distances to reach the nearest step edge.

The experiments are performed in a custom growth chamber at the In-Situ and Resonant Scattering (ISR) beamline at the National Synchrotron Light Source II (NSLS-II). A schematic of the experiment is shown in Figure \ref{Experiment}(a).  Real time X-ray scattering with an energy of 11.3 keV is performed during the growth to extract in-plane and out-of-plane structural information about the growing films.  The measurements are done at the [0 0 1/2] anti-Bragg position  to maximize sensitivity to single unit cell height features.   A broad component of diffuse scattering is visible corresponding to unit-cell height islands on the surface. This image corresponds to 5 ML of deposition, which results in a relatively coarse array of islands. The scattering forms a nearly perfect circle with a radius of $\approx$0.2 nm$^{-1}$, indicating that islands are isotropically arranged on the surface with a mean spacing of $\approx$30 nm.   In our experiments, we are sensitive in-plane diffuse scattering intensity out to about 3 nm$^{-1}$, corresponding to length scales on the order of 2 nm. That is, we can access the whole range of length scales over which the two-dimensional islands form, from the scale of newly-arrived particles in individual laser pulses through the aggregation, coarsening, and coalescence stages of monolayer growth as they occur.

Two growth processes have been developed for these experiments, which we refer to as e-PLD and th-PLD. The energetic process uses parameters similar to previous studies of the homoepitaxy of SrTiO$_3$.\cite{Eres_2011} A background gas of 2 mTorr O$_2$ is used to ensure proper oxidation of the growing surface. The laser fluence on the target is 3 J/cm$^2$, which produces a deposition rate of 18 to 22 pulses per monolayer. The substrate temperature is  600$^\circ$C which results in a smooth starting surface and extended layer-by-layer growth.  For the thermalized process,\cite{Shin_2007} 300 mTorr of  Helium buffer gas is added along with the Oxygen.   Time-of-flight curves obtained with a Langmuir probe placed 8.7 cm from the laser target for each process are shown in Figure \ref{Experiment}(c) and (d).\cite{Doggett_2009,supplemental_citation}  As the helium background pressure is increased, the fast component shown in Figure \ref{Experiment}(c) becomes attenuated and it is completely replaced with a much slower component at 300 mTorr, as shown in Figure \ref{Experiment}(d).   As a result, particle kinetic energies are reduced from $\approx$ 100 eV/atom for e-PLD to $\approx$ 0.5 eV/atom for th-PLD. 

%It should be noted that both of these processes fall into the PLD regime. Surface diffusion coefficients for STO were measured experimentally by Ferguson et al.\cite{Ferguson_2009}. At 600$^\circ$C, the average time for a particle to diffuse one lattice constant is 60 milliseconds, which is much larger that the width of the Thermal process. Therefore we expect a similar island nucleation density after the first pulse for both processes.

%The basic idea of these plots is that the data should fall between the two limiting cases.
%The surprising result is that both processes are similar, implying even the thermal process has a fast relaxation mechanism which is not energetic in nature.

Surface X-ray scattering can be divided into two contributions: (i) the specular reflection, which is only sensitive to the layer coverages, and (ii) the diffuse intensity, which is additionally sensitive to in-plane structure. Figure \ref{Spec_Diff}(a) shows both specular and diffuse intensity data during e-PLD growth of SrTiO$_3$ up to a final thickness of 4.5 unit cells.  The top panel is the specular intensity, which oscillates with each layer, indicating layer-by-layer growth. A defining characteristic of PLD is that the jumps in the specular intensity coincide with the arrival of the plume due to a sudden change in the layer coverages. After each pulse, there is a slow evolution of the intensity  due to interlayer transport as monomers descend and attach to step-edges from above.  Comparable results for th-PLD are shown in Fig. \ref{Spec_Diff}(b). The specular intensity is very similar aside from a slightly reduced intensity at the peaks of the oscillations. The lower panels in Figs. \ref{Spec_Diff} (a) and (b) show diffuse scattering profiles versus deposited thickness.  They reveal an interesting difference that e-PLD exhibits a broad distribution centered at  $\approx$1 nm$^{-1}$, which is absent from the th-PLD profiles. This peak position corresponds to a length scale of  $\approx$5 nm.  As we discuss below, the length scale of this broad profile can be as small as 2.5 nm for shorter dwell times.  

 In order to gain further insight into this difference and it's origin, Fig. \ref{Dwell}(a,b) shows diffuse scattering profiles for various dwell times at the same coverage of $\theta = 0.4$ ML. For these plots, we performed a circular average of the diffuse intensity at each  $Q_r$ where $Q_r = \sqrt{Q_x^2 + Q_y^2}$ in order to improve the counting statistics.   Background corresponding to scattering from the substrate before the deposition was also subtracted, as described in the Supplemental Materials.\cite{supplemental_citation} There is a clear difference in the diffuse scattering profiles, both as a  function of the dwell time and also between e-PLD and th-PLD.  In Fig. \ref{Dwell}(c), we plot the length scale $\ell$ derived from the peak position of the diffuse scattering, $Q_{\mathrm{r,max}}$.   Comparison of the curves for energetic and thermalized deposition reveals the remarkable result that the thermal coarsening effect is inhibited in e-PLD, which implies that the nucleation density is greatly increased by energetic effects.  For example, comparing the curves for 6 sec. dwell time in Fig. \ref{Dwell}(c), we see that  the length scale for e-PLD is about a factor of three smaller than for th-PLD over the entire range of coverages. For shorter dwell times, where thermal surface diffusion and relaxation are less dominant in both cases, we observe that the length scales become progressively smaller, even approaching the diffusion-less limit of $\ell\approx$ 2 nm that we deduced from the pulse intensity.

% Figure 4 - Specular Intensity Jump Analysis
\begin{figure}[ht!]
\includegraphics[scale=0.5]{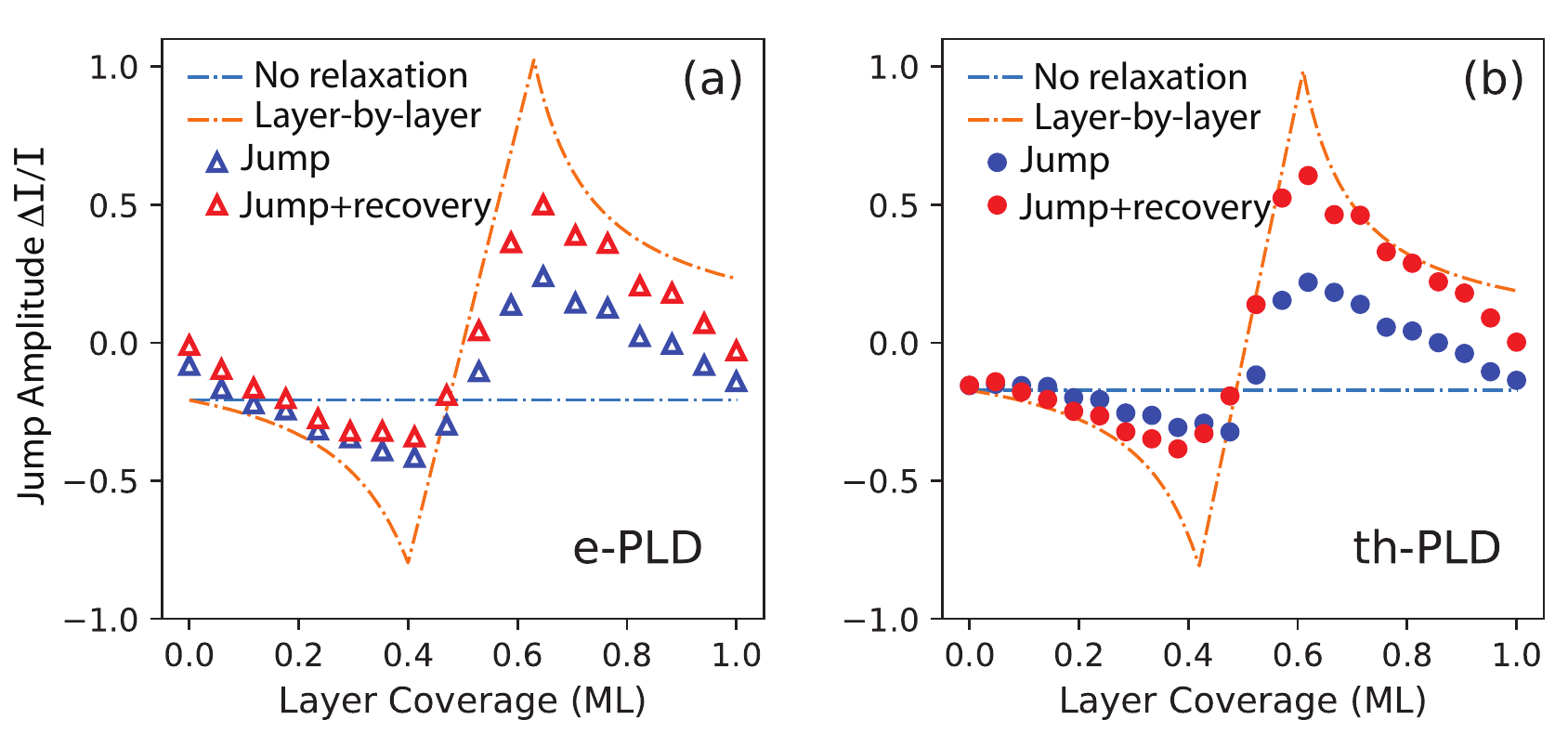}
\caption{\label{Jump} Analysis of the transient jump in the specular intensity following each laser pulse for (a) e-PLD and (b) th-PLD. Blue symbols labelled ``Jump'' represent the immediate normalized intensity change, while the red symbols labelled ``Jump + recovery'' represent the total normalized intensity change at the end of the 6 s dwell time. The dash-dotted lines are the two limiting cases of random deposition with no inter-layer relaxation (blue) and perfect layer-by-layer relaxation (orange).}
\end{figure}

Based on the discussion above, the first main conclusion of this study is that the reduced in-plane length scales and coarsening rates for e-PLD are consistent with the island break-up mechanism.\cite{Pomeroy_2002,Willmott_2006}   This process significantly counteracts the effects of thermal surface diffusion, aggregation,  and ripening,  leading to  an island density that can be an order of magnitude higher for e-PLD compared to th-PLD for long dwell times.  However, a net coarsening still occurs for e-PLD, albeit at a slower rate, as shown in Fig. \ref{Dwell}(c), and it continues during the first few monolayers of growth, as observed in Fig. \ref{Spec_Diff}(a).

The specular intensity can be expressed as a simple function of the layer coverages.  Assuming layer coverages $\theta_n(t)$ for layers $n=1 ~\rm{to}~ N$:

\begin{equation}\label{Eq:Spec_Intensity}
I(t) \propto |F(Q)|^2 \left\vert \frac{1}{1-e^{iQa}} + \sum_{n=1}^N \theta_n(t) e^{-i Qna} \right\vert ^2,
\end{equation}

\noindent where $F(Q)$ is the scattering amplitude of a single unit-cell high layer and $Q$ is the magnitude of the scattering vector, which is oriented along the surface normal for the specular reflection.

Eq. \ref{Eq:Spec_Intensity} can be used to formulate several simple growth models for the intensity at the anti-Bragg position. In one model called the impulse approximation, it is assumed that monomers arrive at random positions on the surface during the pulse and then remain in the same layer. This model predicts a drop in specular intensity $I$ following each laser pulse given by  $\Delta I/I = -4 \sigma (1-\sigma)$.\cite{Fleet_2005}  The opposite limiting case is for perfect layer-by-layer (LBL) growth, which is described by   $\Delta I/I = -4 \sigma ((1-\sigma)-2\theta)/(1-2\theta)^2$.  Note that the impulse model predicts that the intensity jump is always negative, so that in the absence of any relaxation the intensity should continuously decrease.  This is clearly inconsistent with experimental observations of an oscillating intensity during growth. Therefore, interlayer transport must occur following each laser pulse, and so the interesting question becomes: on what time scale does it occur? In order to address this question,  Fig. \ref{Jump} shows plots of the experimental fast intensity jump along with the total change in the intensity after the dwell time (labelled ``jump + recovery'')  for each pulse during the first monolayer of deposition. The results show that the fast jump amplitude significantly deviates from the impulse model.   This is evidence that there must be a fast relaxation component, a result that was previously deduced for e-PLD.\cite{Fleet_2005}  Interestingly, the fast jump amplitudes are almost the same for th-PLD and e-PLD, which is remarkable since previous work emphasized the possible role of high kinetic energies ($E_k>10$ eV) in promoting interlayer transport.\cite{Tischler_2006, Fleet_2005}  In contrast, our results show that the high kinetic energy of the incident particles in e-PLD does not  enhance interlayer transport relative to the more modest energies present in th-PLD.  The data for the intensity change after the dwell time is observed to be significantly closer to the LBL model, which indicates that thermal relaxation also plays a significant role, although it occurs on a much slower timescale. However, there is still a significant deviation from LBL near the middle of the growth cycle, which is related to the fact that the specular intensity $I$  does not disappear completely near $\theta$=0.5 ML.  This effect is likely due to imperfections in the structure of the deposited layer.

The second main conclusion of this study is that fast interlayer transport occurs for th-PLD as well as e-PLD.  We note that based on our Langmuir probe studies that the temperature of incidence particles in th-PLD may be as high as 4000 K.  In this range, ``hot'' particles incident in normal incidence  can scatter nearly elastically from surface atoms, leading to a significant velocity in the plane of the surface.  For example, computational results by Gao et al. for Pd deposition on MgO showed that Pd arriving near Mg surface atoms can recoil with a significant lateral kinetic energy without desorbing.\cite{Gao_2012}  Particles incident with 0.4 eV kinetic energy were found to travel up to 2 nm from their impact site.   Our results indicate that partially thermalized particles in SrTiO$_3$ th-PLD travel up to $\approx$10 nm during growth. 

%There may be other contributions factors, such as the possibility that the charge state of the incident particle can affect the barrier to diffusion.\cite{Hong_2013} 

 In conclusion, our results suggest that  incident particles with kinetic energies  $0.1<E_k<10$ eV can migrate via transient-enhanced surface diffusion.  Higher kinetic energies can break chemical bonds, leading to phenomena such as island breakup. These mechanisms should be included in models of epitaxial growth by hyperthermal and energetic deposition methods, and they can be exploited to engineer thin films and multilayers with improved properties.

The authors acknowledge the contributions of Christie Nelson. This material is based on work that was supported by the National Science Foundation under grant No. DMR-1506930.   JU and RH were partially supported by the U.S. Department of Energy (DOE) Office of Science under Grant No. DE-SC0017802. This research used the 4-ID beamline of the National Synchrotron Light Source II, a U.S. DOE Office of Science User Facility operated for the DOE Office of Science by Brookhaven National Laboratory under Contract No. DE-SC0012704.

%\noindent

%\begin{equation}\label{Eq:Fsphere}
%F_{sphere}(\vec{Q},R) \propto  \frac{4}{3}\pi R^3  \left[ \frac{3(\vec{Q}R\cos \vec{Q}R - \sin \vec{Q}R)}{(\vec{Q}R)^3} \right]
%\end{equation}

%\begin{table*}
%\caption{\label{table1} Calculation of Nanocluster Density. $L$ = 150 mm for all processes.}
%\begin{ruledtabular}
%\begin{tabular}{cccccccccc}
%
%Process & \multicolumn{2}{c}{Argon} & \multicolumn{2}{c}{Helium} & Power & Diameter & Mass &  \multicolumn{2}{c}{Density} \\
%              &Mass Flow      &Pressure   &Mass Flow     &Pressure   \\
%              & (sccm) & (Torr) & (sccm) & (Torr) & (W) & (nm) & (amu) & (amu/nm$^3$) & (g/cm$^3$)\\
%\hline
%WSi$_x$ - I & 20 & 0.243$^a$ & 110 & 0.394$^a$ & 25 & 2.15 & 4.63$\times$10$^4$ & 8.87$\times$10$^3$ & 14.7 $\pm$ 1.8\\
%WSi$_x$ - II & 50 & 0.550~ & 0 & 0~ & 50 & 4.1 & 3.20$\times$10$^5$ &8.67$\times$10$^3$ & 14.3 $\pm$  4.0\\
%WSi$_x$ - III & 50& 0.550~ & 0 & 0~ & 25 & 5.9 & 8.21$\times$10$^5$ & 7.90$\times$10$^3$ & 12.3 $\pm$  6.3\\
%Copper & 100 & 1.000~ & 0 & 0~ & 100 & 4.9 & 3.01$\times$10$^5$ & 5.02$\times$10$^3$ & 8.33 $\pm$ 1.2\\
%
%\end{tabular}
%\end{ruledtabular}
%\begin{tabular*}{0.60\textwidth}{@{\extracolsep{\fill}}cccc}
%$^a$ Partial pressure. 
%\end{tabular*}
%\end{table*}

% Create the reference section using BibTeX:
\bibliography{PLD_STO_Energetic_vs_Thermal_V7}

\end{document}